\documentclass[12pt]{iopart}
\newcommand{\la}{\left\langle}
\newcommand{\ra}{\right\rangle}
\begin{document}
\jl{3}
\title[Effective Interactions in Colloidal Suspensions]
{Effective Interactions and Volume Energies
in Charge-Stabilized Colloidal Suspensions}

\author{A R Denton\footnote[1]{e-mail address: alan.denton@acadiau.ca}}

\address{Department of Physics, Acadia University, 
Wolfville, NS, Canada B0P 1X0}

\begin{abstract}
Charge-stabilized colloidal suspensions can be conveniently described by 
formally reducing the macroion-microion mixture to an equivalent
one-component system of pseudo-particles.  
Within this scheme, the utility of a linear response approximation for 
deriving effective interparticle interactions has been demonstrated 
[M.~J.~Grimson and M.~Silbert, {\it Mol.~Phys.}~{\bf 74}, 397 (1991)].
Here the response approach is extended to suspensions of 
{\it finite-sized} macroions and used to derive explicit expressions for 
(1) an effective electrostatic pair interaction between pseudo-macroions and 
(2) an associated volume energy that contributes to the total free energy.
The derivation recovers precisely the form of the DLVO screened-Coulomb 
effective pair interaction for spherical macroions and makes manifest 
the important influence of the volume energy on thermodynamic properties 
of deionized suspensions.
Excluded volume corrections are implicitly incorporated through a
natural modification of the inverse screening length.
By including {\it nonlinear} response of counterions to macroions, 
the theory may be generalized to systematically investigate effective 
{\it many-body} interactions.
\end{abstract}

\bigskip
\bigskip

\pacs{82.70.Dd, 83.70.Hq, 05.20.Jj, 05.70.-a}
\submitted
\maketitle


\section{Introduction}

Charge-stabilized colloidal suspensions, composed of charged macroions 
($1-1000$ nm in diameter) and microions (counterions and salt ions) 
suspended by Brownian motion in a molecular fluid, occur in a variety of 
natural and engineered forms~\cite{Hunter}.  Common examples include 
clay minerals,  paints, inks, and detergents (micellar solutions), in which
repulsive electrostatic interactions promote stability against coagulation
induced by van der Waals attractive forces.
Somewhat more exotic are suspensions of synthetic latex or silica spheres 
whose near monodispersity facilitates self-assembly into crystalline 
lattices~\cite{Pusey}. 
Aside from serving as well-characterized models for fundamental study, 
synthetic colloids exhibit unique optical properties that underly several 
emerging technological applications, such as nanosecond optical 
switches~\cite{Asher} and photonic band gap materials~\cite{Vos}.  

Accurate prediction of the physical properties of colloidal matter 
relies on a fundamental understanding of interparticle interactions.  
The first quantitative account of electrostatic interactions was achieved
by Derjaguin, Landau, Verwey, and Overbeek (DLVO)~\cite{DLVO}.  Based on
the Poisson-Boltzmann equation for the electrostatic potential~\cite{Hunter},  
the DLVO theory portrays the bare Coulomb interactions between macroions 
as effectively screened by a surrounding atmosphere of microions.  
The resulting screened-Coulomb pair potential has been a valuable cornerstone 
of colloid science for half a century.  Nevertheless, experimental evidence 
for apparent long-range attractions between macroions~\cite{Grier} has 
contributed to renewed interest in colloidal interparticle interactions.

An explicit description of the multi-component mixture of macroions, 
counterions, salt ions, and solvent molecules clearly poses a formidable 
challenge.  Consequently, interactions in such complex systems 
are usually treated at the level of {\it effective} interactions.
Tracing out from the partition function statistical degrees of freedom 
of all but a single component reduces the problem to that of an equivalent 
one-component system of ``pseudo-particles" governed by
an effective thermodynamic-state-dependent interaction~\cite{Rowlinson}.
A range of theoretical and numerical methods have been developed to 
investigate effective interactions in charge-stabilized colloids.
Strategies deployed to date include
Poisson-Boltzmann cell models~\cite{Alexander}, Monte Carlo 
simulation~\cite{Stevens,Damico}, {\it ab initio} simulation~\cite{abinitio}, 
and density-functional theory~\cite{vRH,vRDH,Graf,Goulding}.
Recently, Silbert and coworkers~\cite{Silbert1,Silbert2} proposed an
approach motivated by analogies between charged colloids and metals.  
With the correspondences (counterion $\leftrightarrow$ electron) and 
(macroion $\leftrightarrow$ metallic ion), the procedure is closely akin
to the pseudopotential theory of metals~\cite{PPT}, which can successfully 
account for thermodynamic properties of simple metals~\cite{Hafner}.
Performing a classical trace over microion degrees of freedom and
describing the electrostatic response of the microions to the macroions 
within second-order perturbation theory, results in an effective interaction
between pseudo-macroions {\it and} an associated volume energy that
contributes to the total free energy.
The importance of including the volume energy in calculating thermodynamic
properties of charge-stabilized colloidal suspensions has been emphasized 
recently by a number of authors~\cite{vRH,vRDH,Graf,Silbert1,DL}.  

The main purpose of this paper is to develop further the response approach
to charge-stabilized colloids, extending it, in particular, to suspensions
of finite-sized macroions.  The proposed extensions enforce exclusion of 
microions from the macroion hard cores and explicitly take into account 
the volume excluded by the macroions to the microions.  
A second goal of the paper is to establish a framework for generalizing 
the theory to include nonlinear response to allow investigation of effective 
{\it many-body} interactions.  In the next section, following a brief review 
of the response approach, the extensions are outlined.  Section 3 presents the 
main results -- obtained within a linear response approximation -- for 
an effective pair potential acting between pseudo-macroions and an associated 
volume energy, both of which consistently incorporate excluded volume effects.  
Finally in Sec.~4, we discuss implications of the results for thermodynamic 
properties of charge-stabilized colloids and prospects for generalizing 
the theory beyond linear response.

\section{Theory}

\subsection{The Model}

The theory described below is based on the ``primitive" model, wherein 
for simplicity the solvent is treated as a uniform dielectric continuum.
To simplify notation, it is furthermore assumed that counterions are the 
only microions present (deionized suspension).  The general case of
finite salt concentration will be addressed elsewhere~\cite{Denton}.
The model system then consists of $N_{\rm m}$ charged hard-sphere macroions
of diameter $\sigma$ and charge $-Ze$ ($e$ being the elementary charge) 
and $N_{\rm c}$ point counterions of charge $ze$ suspended in a uniform 
fluid medium characterized entirely by a dielectric constant $\epsilon$. 
Each macroion is assumed to carry a fixed charge, 
uniformly distributed over its surface.  Charge fluctuations are thus
implicitly ignored in the model.
Described statistically by a canonical ensemble, the system occupies a fixed 
total volume $V$ at temperature $T$.  For a given number of macroions, 
global charge neutrality constrains the number of counterions by the 
condition $zN_{\rm c}=ZN_{\rm m}$.  

Denoting macroion and counterion positions by ${\bf R}_i$ and ${\bf r}_i$, 
respectively, the full Hamiltonian of the system can be expressed in the form
\begin{equation}
H~=~H_{\rm m} + H_{\rm c} + H_{\rm mc},
\label{H}
\end{equation}
with the individual terms to be specified below.
The first term on the right side of Eq.~(\ref{H}) is the bare
macroion Hamiltonian, given by
\begin{equation}
H_{\rm m}~=~K_{\rm m} + {1\over{2}}\sum_{{i,j=1}\atop{(i\neq j)}}^{N_{\rm m}}
\Bigl[v_{\rm HS}(|{\bf R}_i-{\bf R}_j|)+v_{\rm mm}(|{\bf R}_i-{\bf R}_j|)\Bigr],
\label{Hm}
\end{equation}
where $K_{\rm m}$ is the kinetic energy of the macroions, 
$v_{\rm HS}(|{\bf R}_i-{\bf R}_j|)$ is a hard-sphere pair interaction 
between the macroion cores, and $v_{\rm mm}(r)=Z^2e^2/\epsilon r$ is the 
bare Coulomb interaction between a pair of macroions whose centres are
separated by a distance $r>\sigma$.
The second term in Eq.~(\ref{H}) is a counterion Hamiltonian, taking the form
\begin{equation}
H_{\rm c}~=~K_{\rm c} + {1\over{2}}\sum_{{i,j=1}\atop{(i\neq j)}}^{N_{\rm c}}
v_{\rm cc}(|{\bf r}_i-{\bf r}_j|) +
\sum_{i=1}^{N_{\rm c}}\sum_{j=1}^{N_{\rm m}} v_{\rm HS}(|{\bf r}_i-{\bf R}_j|),
\label{Hc}
\end{equation}
where $K_{\rm c}$ is the counterion kinetic energy, 
$v_{\rm cc}(r)=z^2e^2/\epsilon r$ is the Coulomb interaction between 
a pair of counterions, and $v_{\rm HS}(|{\bf r}_i-{\bf R}_j|)$ is the
hard-sphere interaction between a point counterion and a macroion core.
Finally, the third term in Eq.~(\ref{H}) is the electrostatic 
interaction energy between the macroions and counterions:
\begin{equation}
H_{\rm mc}~=~\sum_{i=1}^{N_{\rm c}}\sum_{j=1}^{N_{\rm m}} 
v_{\rm mc}(|{\bf r}_i-{\bf R}_j|),
\label{Hmc1}
\end{equation}
where $v_{\rm mc}(r)$ denotes the macroion-counterion electrostatic 
pair interaction.  Outside the macroion core radius, $v_{\rm mc}(r)$ 
has the Coulomb form.  Inside the core, however, $v_{\rm mc}(r)$ 
is not uniquely defined.  Thus, following van Roij and Hansen~\cite{vRH}, 
we {\it choose} $v_{\rm mc}(r)$ to be a constant for $r<\sigma$ and take
\begin{equation}
v_{\rm mc}(r)~=~\left\{ \begin{array}
{l@{\quad\quad}l}
\frac{\displaystyle -Zze^2}{\displaystyle \epsilon r}, & r>\sigma/2 \\
\frac{\displaystyle -Zze^2}{\displaystyle \epsilon\sigma/2}\alpha, & r<\sigma/2,
\end{array} \right.
\label{vmcr}
\end{equation}
where the parameter $\alpha$ will be specified (Sec.~3.1) to ensure that 
the counterion density vanishes within the core, as physically it must.

\subsection{Reduction to an Equivalent One-Component System}

Having defined the Hamiltonian, we now turn to a statistical mechanical 
description, with the ultimate aim of calculating the free energy of the 
system.  The partition function is given by
\begin{equation}
{\cal Z}_N~=~\la\la\exp(-H/k_{\rm B}T)\ra_{\rm c}\ra_{\rm m},
\label{part1}
\end{equation}
the angular brackets symbolizing classical traces over macroion or 
counterion degrees of freedom.
Following standard treatments developed first in the context of simple 
metals~\cite{Hafner,HM}, we proceed by reducing the two-component mixture 
of macroions and counterions to an equivalent one-component system.  
The reduction is achieved by performing a restricted trace over counterion 
coordinates, while keeping the macroion coordinates fixed.  
Thus, without approximation in this purely classical system,
\begin{equation}
{\cal Z}_N~=~\la\exp(-H_{\rm eff}/k_{\rm B}T)\ra_{\rm m},
\label{part2}
\end{equation}
where $H_{\rm eff}\equiv H_{\rm m}+F_{\rm c}$ is the effective Hamiltonian of 
a one-component system of pseudo-macroions, and where
\begin{equation}
F_{\rm c}~\equiv~-k_{\rm B}T\ln\la 
\exp\Bigl[-(H_{\rm c}+H_{\rm mc})/k_{\rm B}T\Bigr]\ra_{\rm c}
\label{Fc1}
\end{equation}
may be physically interpreted as the free energy of a nonuniform gas of 
counterions in the presence of macroions fixed at positions ${\bf R}_i$.
In general, the counterion free energy is a complicated many-body function of 
the macroion positions.  Progress can be made, however, by formally 
adding to and substracting from $H$ a term, $E_{\rm b}$, 
representing the energy of a uniform compensating negative background.
Then $F_{\rm c}$ may be expressed in the form
\begin{equation}
F_{\rm c}~=~-k_{\rm B}T\ln\la\exp\Bigl[-(H_{\rm c}' +
H_{\rm mc}')/k_{\rm B}T\Bigr]\ra_{\rm c},
\label{Fc2}
\end{equation}
where $H_{\rm c}'\equiv H_{\rm c}+E_{\rm b}$ and
$H_{\rm mc}'\equiv H_{\rm mc}-E_{\rm b}$.  
The advantage of this manipulation is that $H_{\rm c}'$ is simply 
the Hamiltonian of a classical one-component plasma (OCP) of counterions
in the presence of {\it neutral} hard-sphere macroions.

Now the counterions, being excluded by the hard cores of the macroions, 
occupy the {\it free} volume $V_{\rm f}\equiv V-N_{\rm m}(\pi/6)\sigma^3$, 
{\it i.e.}, the volume not occupied by the macroion cores.  
The average {\it effective} density of counterions is therefore given by 
$n_{\rm c}=N_{\rm c}/V_{\rm f}=n^{(o)}_{\rm c}/(1-\eta)$, where
$n^{(o)}_{\rm c}\equiv N_{\rm c}/V$ is the {\it nominal} counterion density 
and $\eta\equiv (V-V_{\rm f})/V$ is the macroion volume fraction.
An important question now arises.  What volume should the background occupy?
In order that $H_{\rm c}'$ truly be the Hamiltonian of an OCP, the background
and counterions clearly must occupy the same volume.  In fact, were the 
background to occupy a different volume ({\it e.g.}, the total volume $V$) 
then the effective Hamiltonian would contain terms that are formally infinite 
(see below), although identically cancelling, associated with the 
long-range Coulomb interaction.  Thus, the background is taken to be excluded 
-- along with the counterions -- from the macroion cores, its density equaling 
the effective counterion density $n_{\rm c}$.

The background energy is then given explicitly by~\cite{HM}
\begin{equation}
E_{\rm b}~=~-{1\over{2}}n_{\rm c}^2\int_{V_{\rm f}}
{\rm d}{\bf r} \int_{V_{\rm f}}{\rm d}{\bf r}'{{z^2e^2}\over{\epsilon 
|{\bf r}-{\bf r}'|}}~=~-{1\over{2}}N_{\rm c}n_{\rm c}\hat v_{\rm cc}(0),
\label{Vb}
\end{equation}
where 
\begin{equation}
\hat v_{\rm cc}(0)~=~\int_{V_{\rm f}}{\rm d}{\bf r}\frac{z^2e^2}{\epsilon r}
~=~\lim_{k\to 0}\Bigl({{4\pi z^2e^2}\over{\epsilon k^2}}\Bigr)
\label{vcc0}
\end{equation}
is the $k \to 0$ limit of the Fourier transform of $v_{\rm cc}(r)$.
The infinity arising from $E_{\rm b}$ will be seen below to be 
formally cancelled by an identical infinity in $H_{\rm mc}$.  

\subsection{Linear Response Approximation}
 
Thus far, the theory is exact, within the primitive model.  The challenge
lies ahead in calculating the counterion free energy [Eq.~(\ref{Fc1})].
One proposed approach~\cite{vRH} invokes density-functional theory to 
approximate $F_{\rm c}$, regarded as a functional of
the counterion density, by expanding in a functional Taylor series about
a uniform counterion OCP.
An alternative strategy~\cite{Silbert1,Silbert2}, inspired by the
pseudopotential theory of metals, is to formally regard $H_{\rm mc}'$ 
as an ``external" potential acting upon a counterion OCP and approximate 
$F_{\rm c}$ by perturbation theory.  
Following the latter strategy\cite{HM}, we write
\begin{equation}
F_{\rm c}~=~F_{\rm OCP}~+~\int_0^1{\rm d}\lambda\la H_{\rm mc}'\ra_{\lambda},
\label{Fc3}
\end{equation}
where 
\begin{equation}
F_{\rm OCP}~=~-k_{\rm B}T\ln\la\exp(-H_{\rm c}'/k_{\rm B}T)
\ra_{\rm c}
\label{FOCP1}
\end{equation}
is the free energy of the reference counterion OCP, occupying a volume 
$V_{\rm f}$, in the presence of neutral hard-sphere macroions.
The integral over $\lambda$ in Eq.~(\ref{Fc3}) physically corresponds to
an adiabatic charging of the macroions from neutral to fully-charged spheres.
The ensemble average $\la~\ra_{\lambda}$ represents an average 
with respect to the distribution function of a system whose macroions 
are identically configured but carry a charge $\lambda Z$.

Further progress is facilitated by expressing $\la H_{\rm mc}'\ra_{\lambda}$ 
in terms of Fourier components of the macroion and counterion densities 
and of the macroion-counterion interaction, according to
\begin{equation}
\fl \la H_{\rm mc}'\ra_{\lambda}~=~
\frac{1}{V_{\rm f}} \sum_{{\bf k}\neq 0} \hat v_{\rm mc}(k) 
\la\hat\rho_{\rm c}({\bf k})\ra_{\lambda} \hat\rho_{\rm m}(-{\bf k}) + 
\frac{1}{V_{\rm f}} \lim_{k\to 0}\Bigl[ \hat v_{\rm mc}(k)
\la\hat\rho_{\rm c}({\bf k})\ra_{\lambda} \hat\rho_{\rm m}(-{\bf k})\Bigr] - 
E_{\rm b}.
\label{Hmc2}
\end{equation}
Evidently $\la H_{\rm mc}'\ra_{\lambda}$ depends through
$\hat\rho_{\rm c}({\bf k})$ upon the
response of the counterions to the macroion charge density.
Regarding the macroion charge as imposing an external potential on the 
counterions, the counterion density may be expressed in the form~\cite{AS}
\begin{eqnarray}
\fl \hat\rho_{\rm c}({\bf k})&=&\chi^{(1)}(k)\hat v_{\rm mc}(k)\hat\rho_{\rm m}
({\bf k}) + \frac{1}{V_{\rm f}}\sum_{\bf q}\chi^{(2)}({\bf q},{\bf k}-{\bf q})
\hat v_{\rm mc}(q)\hat v_{\rm mc}(|{\bf k}-{\bf q}|)\hat\rho_{\rm m}
({\bf q})\hat\rho_{\rm m}({\bf k}-{\bf q}) \nonumber \\
\fl&+& \cdots,
\label{rhoc}
\end{eqnarray}
where $\chi^{(i)}$ is the $i^{\rm th}$ member of a hierarchy of 
response functions of the reference counterion OCP.  
Here, as in ref.~\cite{Silbert1}, we adopt the simplest nontrivial 
approximation and assume that the counterions respond {\it linearly} 
to the macroion charges.  Although its range of validity is uncertain, 
linearization is expected to be justified for sufficiently dilute suspensions 
and weakly charged macroions.
Thus we take
\begin{equation}
\la\hat\rho_{\rm c}({\bf k})\ra_{\lambda}~=~\chi(k)\lambda\hat v_{\rm mc}(k)
\hat\rho_{\rm m}({\bf k}),\qquad k\neq 0,
\label{rhock1}
\end{equation}
where $\chi(k)\equiv \chi^{(1)}(k)$ is the linear response function.
Note that for $k=0$ there is no response, since $\hat\rho_{\rm c}(0)=N_{\rm c}$
is fixed by the number of counterions.
Substituting Eqs.~(\ref{rhock1}) and (\ref{Hmc2}) into Eq.~(\ref{Fc3})
and integrating over $\lambda$, the counterion free energy is given
to second order in the macroion-counterion interaction by
\begin{equation}
\fl F_{\rm c}~=~F_{\rm OCP}~+~\frac{1}{2V_{\rm f}}\sum_{{\bf k}\neq 0}\chi(k)
\left[\hat v_{\rm mc}(k)\right]^2\hat\rho_{\rm m}({\bf k})\hat\rho_{\rm m}
(-{\bf k}) + n_{\rm c}\lim_{k\to 0}\Bigl[N_{\rm m}\hat v_{\rm mc}(k) + 
\frac{N_{\rm c}}{2}\hat v_{\rm cc}(k)\Bigr].
\label{Fc4}
\end{equation}
Correspondingly, the effective Hamiltonian takes the form
\begin{eqnarray}
\fl H_{\rm eff}&=&K_{\rm m}+{1\over{2}}\sum_{{i,j=1}\atop{i\neq j}}^{N_{\rm m}}
v_{\rm HS}(|{\bf R}_i-{\bf R}_j|)+
\frac{1}{2V_{\rm f}}\sum_{\bf k}\hat v_{\rm mm}(k)
\Bigl[\hat\rho_{\rm m}({\bf k})\hat\rho_{\rm m}(-{\bf k})-N_{\rm m}\Bigr] 
\nonumber \\
\fl&+& F_{\rm OCP} + \frac{1}{2V_{\rm f}}\sum_{\bf k}\chi(k)
\left[\hat v_{\rm mc}(k)\right]^2
\hat\rho_{\rm m}({\bf k})\hat\rho_{\rm m}(-{\bf k}) \nonumber \\
\fl&+& n_{\rm c}\lim_{k\to 0}\Bigl[-{{zN_{\rm m}}\over{2Z}}\chi(k)
\left[\hat v_{\rm mc}(k)\right]^2 + N_{\rm m}\hat v_{\rm mc}(k) + 
{{N_{\rm c}}\over{2}}\hat v_{\rm cc}(k)\Bigr].
\label{Heff1}
\end{eqnarray}
Notice, however, that Eq.~(\ref{Heff1}) may be restructured and written 
in the form
\begin{eqnarray}
\fl H_{\rm eff}&=&K_{\rm m}+{1\over{2}}\sum_{{i,j=1}\atop{i\neq j}}^{N_{\rm m}}
v_{\rm HS}(|{\bf R}_i-{\bf R}_j|) +
\frac{1}{2V_{\rm f}}\sum_{\bf k}\hat v_{\rm eff}(k)
\Bigl[\hat\rho_{\rm m}({\bf k})\hat\rho_{\rm m}(-{\bf k})-N_{\rm m}\Bigr] 
+ E_{\rm o} \nonumber\\
\fl&=&K_{\rm m} + {1\over{2}}\sum_{{i,j=1}\atop{i\neq j}}^{N_{\rm m}}
\Bigl[v_{\rm HS}(|{\bf R}_i-{\bf R}_j|) + 
v_{\rm eff}(|{\bf R}_i-{\bf R}_j|)\Bigr] + E_{\rm o},
\label{Heff2}
\end{eqnarray}
where
\begin{equation}
\hat v_{\rm eff}(k)~=~\hat v_{\rm mm}(k) + \hat v_{\rm ind}(k)
\label{veffk}
\end{equation}
may be interpreted as an {\it effective} electrostatic pair potential between 
pseudo-macroions, being the sum of the bare Coulomb potential and an 
{\it induced} potential
\begin{equation}
\hat v_{\rm ind}(k)~=~\chi(k)\left[\hat v_{\rm mc}(k)\right]^2.
\label{vindk1}
\end{equation}
The final term in Eq.~(\ref{Heff2}), 
\begin{equation}
\fl E_{\rm o}~=~F_{\rm OCP} + {{N_{\rm m}}\over{2}}\lim_{r\to 0} v_{\rm ind}(r)
+ n_{\rm c}N_{\rm m}\lim_{k\to 0}\Bigl[-{{z}\over{2Z}}\hat v_{\rm ind}(k)
+ \hat v_{\rm mc}(k) + {{Z}\over{2z}}\hat v_{\rm cc}(k)\Bigr],
\label{Eo1}
\end{equation}
is the {\it volume energy}, which is a natural and inevitable
consequence of the reduction to an equivalent one-component system.  
Although having no explicit dependence on the macroion positions (see below), 
$E_{\rm o}$ evidently depends on the average density of macroions and thus 
can make a significant contribution to the total free energy of the system.
It must be emphasized that the above expressions for the effective pair 
potential and the volume energy are identical to expressions derived from 
the pseudopotential theory of metals~\cite{Hafner,HM,Finnis} if one 
substitutes for $F_{\rm OCP}$ and $\chi(k)$,  respectively, the energy and 
linear response function of the uniform electron gas, and for 
$\hat v_{\rm mc}(k)$ the electron-ion pseudopotential.

To summarize thus far, 
starting from the primitive model of charge-stabilized colloids, formally 
reducing the two-component macroion-counterion mixture to an equivalent
one-component system of pseudo-macroions, and applying a linear response 
approximation to the counterion density, we have obtained expressions for 
both an effective electrostatic pair interaction [Eqs.~(\ref{veffk}) and 
(\ref{vindk1})] and an associated volume energy [Eq.~(\ref{Eo1})].
Practical calculations still require specification of (1) the 
reference OCP free energy $F_{\rm OCP}$, (2) the OCP linear response function 
$\chi(k)$, and (3) the macroion-counterion interaction $\hat v_{\rm mc}(k)$.
Below each of these is considered in turn.

It is important first to note that by associating the hard-sphere part 
of the total macroion-counterion interaction with the counterion Hamiltonian 
[Eq.~(\ref{Hc})] -- necessary, since response theory does not apply to 
hard-sphere interactions -- the reference OCP is confined to the 
free volume between the macroion cores. 
As a consequence, the OCP is not strictly uniform since, in principle, 
the boundary conditions may induce nonuniformity.
Determining the free energy of such a system in general poses a nontrivial 
task.  In practice, however, counterion densities are usually low enough
that the OCP may be assumed to be essentially uniform, except perhaps 
near contact with a macroion surface.

Now, for typical macroion charges and concentrations, the OCP is so 
weakly coupled (unlike its electronic counterpart in metals) that its 
free energy is dominated by the ideal-gas entropic component.  
Therefore, ignoring correlations between counterions~\cite{vRH,Graf,DL}, 
an accurate approximation is
\begin{equation}
\fl F_{\rm OCP}~\simeq~k_{\rm B}TN_{\rm c}\Bigl[\ln(n_{\rm c}\Lambda^3)-1\Bigr]
~=~k_{\rm B}TN_{\rm c}\Bigl[\ln\Bigl(\frac{(Z/z)n_{\rm m}\Lambda^3}{1-\eta}
\Bigr)-1 \Bigr],
\label{FOCP2}
\end{equation}
where $\Lambda$ is the thermal de Broglie wavelength and
$n_{\rm m}\equiv N_{\rm m}/V$ is the average number density of macroions, 
the last equality following from the constraint of global charge neutrality.

The linear response function is directly related to 
the corresponding static structure factor $S(k)$ via
\begin{equation}
\chi(k)~=~-\beta n_{\rm c}S(k)~=~-\frac{\beta n_{\rm c}}{1-n_{\rm c}\hat c(k)},
\label{chi1}
\end{equation}
where $\beta\equiv 1/k_{\rm B}T$ and $\hat c(k)$ is the Fourier transform of 
the direct correlation function $c(r)$.  
Specifying $\chi(k)$ is therefore equivalent to specifying $\hat c(k)$.  
For a weakly coupled OCP, a convenient and reasonable approximation for $c(r)$ 
is given by the mean spherical approximation (MSA).  This amounts to setting 
$c(r)$ equal to its asymptotic ($r \to \infty$) limit 
$c(r)\simeq -\beta v_{\rm cc}(r)$ for {\it all} $r$.  As a result, 
\begin{equation}
\hat c(k)~\simeq~-\beta\hat v_{\rm cc}(k)~=~-\frac{4\pi\beta z^2e^2}
{\epsilon k^2}.
\label{MSA}
\end{equation}
Substitution of Eq.~(\ref{MSA}) into Eq.~(\ref{chi1}) then 
yields~\cite{Silbert1}
\begin{equation}
\chi(k)~=~-\frac{\beta n_{\rm c}}{1+\kappa^2/k^2},
\label{chi2}
\end{equation}
where 
\begin{equation}
\kappa\equiv\Bigl(\frac{4\pi n_{\rm c}z^2e^2}{\epsilon k_{\rm B}T}\Bigr)^{1/2}
=\Bigl(\frac{4\pi n^{(o)}_{\rm c}z^2e^2}{(1-\eta)\epsilon k_{\rm B}T}
\Bigr)^{1/2}.
\label{kappa}
\end{equation}
As will be seen below, the parameter $\kappa$ plays the role of
an inverse screening length in the counterion density profile
and in the effective pair interaction.

Finally, specifying the macroion-counterion interaction amounts to 
determining the value of the parameter $\alpha$ in Eq.~(\ref{vmcr}) 
that will ensure a vanishing counterion density inside the macroion cores.  
This in turn requires a calculation of the real-space counterion 
density profile, the details of which are described in the next section.

\section{Results}

\subsection{Counterion Density Profile}

The real-space counterion density profile $\rho_{\rm c}({\bf r})$ 
may be determined from Eqs.~(\ref{vmcr}), (\ref{rhock1}), and (\ref{chi2}).  
First, Fourier transforming Eq.~(\ref{vmcr}) yields
\begin{equation}
\hat v_{\rm mc}(k)~=~-\frac{4\pi Zze^2}{\epsilon k^2}\Bigl[(1-\alpha)
\cos(k\sigma/2)+\alpha\frac{\sin(k\sigma/2)}{k\sigma/2}\Bigr].
\label{vmck1}
\end{equation}
Next, substituting Eqs.~(\ref{chi2}) and (\ref{vmck1}) into 
Eq.~(\ref{rhock1}) gives the intermediate result
\begin{equation}
\fl\hat \rho_{\rm c}({\bf k})~=~\frac{Z}{z}\Bigl(\frac{\kappa^2}
{k^2+\kappa^2}\Bigr)\Bigl[(1-\alpha)\cos(k\sigma/2)+\alpha
\frac{\sin(k\sigma/2)}{k\sigma/2}\Bigr]
~\sum_{{\bf R}}\exp(i{\bf k}\cdot{\bf R}),
\label{rhock2}
\end{equation}
where the sum is over the positions ${\bf R}$ of the macroions.
For simplicity, we consider the density profile around a single macroion
located at the origin (${\bf R}=0$), 
assuming all other macroions to be far away ($\kappa R\gg 1$). 
This is equivalent to retaining only the ${\bf R}=0$ term in the summation, 
in which case $\hat\rho_{\rm c}({\bf k})$ is a function only of $k$
and $\rho_{\rm c}({\bf r})$ is a function only of the radial distance $r$.  
Now inverse Fourier transforming Eq.~(\ref{rhock2}) yields
\begin{equation}
\fl\rho_{\rm c}(r)~=~\left\{ \begin{array}
{l@{\quad\quad}l}
\frac{\displaystyle Z}{\displaystyle z}\frac{\displaystyle \kappa^2}
{\displaystyle 4\pi}\Bigl[(1-\alpha)
\cosh(\kappa\sigma/2)+\alpha\frac{\displaystyle \sinh(\kappa\sigma/2)}
{\displaystyle \kappa\sigma/2}\Bigr]
~\frac{\displaystyle \exp(-\kappa r)}{\displaystyle r}, & r>\sigma/2 \\
\frac{\displaystyle Z}{\displaystyle z}\frac{\displaystyle \kappa^2}
{\displaystyle 4\pi}
\Bigl(-1+\alpha+\frac{\displaystyle \alpha}{\displaystyle \kappa\sigma/2}\Bigr) 
{\displaystyle \e^{-\kappa\sigma/2}}~
\frac{\displaystyle \sinh(\kappa r)}{\displaystyle r}, & r<\sigma/2. 
\end{array} \right.
\label{rhocr1}
\end{equation}
Vanishing of $\rho_{\rm c}(r)$ for $r<\sigma/2$ is evidently ensured by setting
\begin{equation}
\alpha~=~\frac{\kappa\sigma/2}{1+\kappa\sigma/2}.
\label{alpha}
\end{equation}
Finally, substituting this expression for $\alpha$ back into Eq.~(\ref{rhocr1}) 
gives the result
\begin{equation}
\rho_{\rm c}(r)~=~\frac{Z}{z}\frac{\kappa^2}{4\pi}
\Bigl(\frac{\e^{\kappa\sigma/2}}{1+\kappa\sigma/2}\Bigr) 
\frac{\e^{-\kappa r}}{r}, \qquad r>\sigma/2,
\label{rhocr2}
\end{equation}
which is automatically normalized to the correct number of counterions 
per macroion ($Z/z$).
This expression for the counterion density profile around a single macroion is 
recognized to be of precisely the same form as the Debye-H\"uckel expression
for the density of electrolyte ions around a macroion~\cite{Hunter}, where
$\kappa$ is the inverse Debye screening length.
A notable distinction lies, however, in the definition of $\kappa$.  
Whereas our $\kappa$ [Eq.~(\ref{kappa})] depends on the average 
{\it effective} counterion density $n_{\rm c}$ in the volume unoccupied by 
macroions, the Debye-H\"uckel $\kappa$ depends rather on the 
{\it nominal} bulk density of electrolyte ions.
The importance of redefining the usual $\kappa$ in this way -- a result that
emerges naturally from the response approach -- has been stressed also by 
Russel and coworkers~\cite{Russel}.

In passing, two remarks are in order.  First, determining the counterion 
density profile in the presence of two or more closely spaced macroions 
will evidently require a more general form for the macroion-counterion 
core interaction than the simple constant chosen in Eq.~(\ref{vmcr}).  
Second, it may be instructive to compare the macroion-counterion interaction 
with its metallic counterpart, the electron-ion pseudopotential.  
A popular and successful form of the latter is the empty-hole 
pseudopotential~\cite{Ashcroft}, which is Coulombic at long range but
precisely zero inside a certain core radius.
In contrast, setting $v_{\rm mc}(r)$ to zero for $r<\sigma/2$ would result 
in a nonvanishing counterion density inside the macroion cores.
A distinction between the metallic and colloidal cases lies in the
fact that, while counterions are strictly excluded from macroion cores, 
electrons may at least partially penetrate metallic ion cores.

\subsection{Effective Pair Interaction and Volume Energy}

We are now in a position to derive the main results of the paper.
Considering first the effective electrostatic pair interaction between 
pseudo-macroions, we proceed by substituting Eq.~(\ref{alpha}) 
into Eq.~(\ref{vmck1}), obtaining
\begin{equation}
\hat v_{\rm mc}(k)~=~-\frac{4\pi Zze^2}{\epsilon k^2}\Bigl(\frac{1}
{1+\kappa\sigma/2}\Bigr)\Bigl[\cos(k\sigma/2)+\kappa\frac{\sin(k\sigma/2)}{k}
\Bigr].
\label{vmck2}
\end{equation}
Next substituting Eqs.~(\ref{chi2}) and (\ref{vmck2}) into Eq.~(\ref{vindk1})
yields 
\begin{equation}
\fl \hat v_{\rm ind}(k)~=~-\frac{2\pi Z^2e^2}{\epsilon k^2}\Bigl(\frac{1}
{1+\kappa\sigma/2}\Bigr)^2 \Bigl(\frac{\kappa^2}{k^2+\kappa^2}\Bigr)
\Bigl[1+\cos(k\sigma) + 2\kappa\frac{\sin(k\sigma)}{k} + 
\kappa^2\frac{1-\cos(k\sigma)}{k^2}\Bigr].
\label{vindk2}
\end{equation}
Fourier transformation of Eq.~(\ref{vindk2}) is a straightforward, if tedious,  
calculation, with the result 
\begin{equation}
v_{\rm ind}(r)~=~\left\{ \begin{array}
{l@{\quad\quad}l}
\frac{\displaystyle Z^2e^2}{\displaystyle \epsilon}\Bigl(
\frac{\displaystyle \e^{\kappa\sigma/2}}
{\displaystyle 1+\kappa\sigma/2}\Bigr)^2~
\frac{\displaystyle \e^{-\kappa r}}{\displaystyle r} - 
\frac{\displaystyle Z^2e^2}{\displaystyle \epsilon r}, & r>\sigma \\
-\frac{\displaystyle Z^2e^2}{\displaystyle 2\epsilon r}\Bigl(
\frac{\displaystyle 1}{\displaystyle 1+\kappa\sigma/2}\Bigr)^2
\Bigl[(2+\kappa\sigma)\kappa r-\frac{1}{2} \kappa^2 r^2\Bigr], & r<\sigma.
\end{array} \right.
\label{vindr}
\end{equation}
Finally, substituting Eq.~(\ref{vindr}) into the Fourier transform of
Eq.~(\ref{veffk}), we obtain an explicit expression for the real-space
form of the effective electrostatic pair potential: 
\begin{equation}
v_{\rm eff}(r)~=~v_{\rm mm}(r) + v_{\rm ind}(r)~=~
\frac{Z^2e^2}{\epsilon}\Bigl(\frac{\e^{\kappa\sigma/2}}
{1+\kappa\sigma/2}\Bigr)^2~\frac{\e^{-\kappa r}}{r}, \qquad r>\sigma. 
\label{veffr}
\end{equation}
This result is identical in form to the electrostatic part of the familiar 
DLVO effective pair potential~\cite{DLVO}, which is usually derived from 
the Poisson-Boltzmann equation.
The only difference between our potential and the DLVO potential lies in 
the definition of $\kappa$ [Eq.~(\ref{kappa})], which here involves the 
{\it effective} counterion density $n_{\rm c}$.

The volume energy now may be explicitly determined from Eq.~(\ref{Eo1}).  
It follows immediately from Eq.~(\ref{vindr}) that
\begin{equation}
\lim_{r\to 0} v_{\rm ind}(r)~=~-\frac{Z^2e^2}{\epsilon}\frac{\kappa}
{1+\kappa\sigma/2},
\label{vindr0}
\end{equation}
from Eq.~(\ref{vindk2}) that
\begin{equation}
\fl \lim_{k\to 0} \hat v_{\rm ind}(k)~=~-\Bigl(\frac{Z}{z}\Bigr)^2
\hat v_{\rm cc}(0) + \frac{4\pi Z^2e^2}{\epsilon\kappa^2}+
\frac{\pi Z^2e^2\sigma^2}{\epsilon}\Bigl(\frac{1}{1+\kappa\sigma/2}\Bigr)^2
(1+\frac{2}{3}\kappa\sigma+\frac{1}{12}\kappa^2\sigma^2),
\label{vindk0}
\end{equation}
and from Eq.~(\ref{vmck2}) that
\begin{equation}
\lim_{k\to 0} \hat v_{\rm mc}(k)~=~-\frac{Z}{z}\hat v_{\rm cc}(0) 
+ \frac{\pi Zze^2}{2\epsilon}\Bigl(\frac{1}
{1+\kappa\sigma/2}\Bigr)(1+\frac{1}{6}\kappa\sigma).
\label{vmck0}
\end{equation}
Substituting Eqs.~(\ref{vindr0}), (\ref{vindk0}), and (\ref{vmck0}) 
into Eq.~(\ref{Eo1}), we obtain the following result for the volume energy:
\begin{equation}
E_{\rm o}~=~F_{\rm OCP} - N_{\rm m}\frac{Z^2e^2}{2\epsilon}
\frac{\kappa}{1+\kappa\sigma/2} - N_{\rm m}\frac{Zk_{\rm B}T}{2z}.
\label{Eo2}
\end{equation}
The first term on the right side of Eq.~(\ref{Eo2}) is the OCP free energy, 
discussed in the previous section.  The second term, which depends implicitly
on the macroion density through the parameter $\kappa$, may be given a 
physical interpretation as one half the electrostatic energy associated with
a single pseudo-macroion, composed of a macroion surrounded by its own 
screening cloud of counterions~\cite{vRH,Finnis}.  The final term,  
corresponding to the $k\to 0$ limit in Eq.~(\ref{Eo1}), contributes 
a density-independent constant to the free energy per macroion and hence 
has no influence on thermodynamic phase transitions at zero salt 
concentration.  At finite salt concentration, however, the corresponding
term cannot be ignored~\cite{vRDH,Graf,Denton}.

\section{Discussion and Conclusions}

It is important to point out some limitations of the theory and 
the results presented above.  
First of all, the assumption of linear response of the counterions 
is strictly valid only for dilute suspensions of weakly charged macroions.
Whether the linear response approximation remains valid at higher
concentrations -- in particular, concentrations for which excluded volume 
effects begin to play a role -- is an interesting and open question.
In this regard, it may be worth noting that Poisson-Boltzmann cell model 
calculations~\cite{Alexander} and {\it ab initio} simulations~\cite{abinitio} 
do support the general form of the screened-Coulomb pair potential 
at appreciable concentrations, albeit with renormalized DLVO parameters.
Secondly, by considering only electrostatic and hard steric interactions, 
and ignoring short-range interactions between counterions and macroion 
surfaces, the theory cannot address the possibility of condensation of 
counterions onto the macroions and the consequences for effective 
macroion charges.  Finally, no account is taken of correlations 
between charge fluctuations, either on the macroion surfaces or in the 
density distributions of counterions surrounding neighbouring macroions.  
For spherical macroions, Monte Carlo simulations and cell model 
calculations~\cite{Jonsson} suggest that such correlations make only a small 
contribution to the total free energy, at least at low salt concentrations.  
On the other hand, for rod-like macroions, correlated charge fluctuations 
may play a more significant role~\cite{Liu}.

In summary, by applying a linear response approximation to the counterions 
surrounding charged monodisperse hard-sphere macroions in a 
colloidal suspension, we have derived two main results, namely an 
effective electrostatic pair interaction $v_{\rm eff}(r)$ [Eq.~(\ref{veffr})] 
and an associated volume energy $E_{\rm o}$ [Eq.~(\ref{Eo2})].  
The total free energy of the system is ultimately the sum of the volume energy 
-- whose physical origins are the counterion entropy and the 
macroion-counterion interaction energy -- and the free energy of the 
equivalent one-component system of pseudo-macroions interacting via 
their effective pair interaction.  Our expression for 
$v_{\rm eff}(r)$ confirms that a linear response approximation, combined 
with the MSA for the response function, yields the familiar DLVO form of 
pair potential for spherical macroions, prefactors included~\cite{note}.  
This is not surprising, given that the DLVO potential also may be derived 
by linearizing the Poisson-Boltzmann equation.  At the same time, 
however, our derivation indicates how excluded volume corrections may be 
incorporated through the density dependence of the inverse screening length 
$\kappa$, by substituting for the nominal counterion density $n^{(o)}_{\rm c}$
the {\it effective} density $n_{\rm c}$ of counterions occupying the 
free volume between macroion cores.  

Our expression for the volume energy, which quantitatively exhibits the 
dependence on the macroion density, confirms the necessity of including 
$E_{\rm o}$ in calculating thermodynamic properties from the 
free energy~\cite{Silbert1}.
In particular, the phase behaviour of deionized suspensions of 
highly charged macroions have been shown to depend sensitively on
the volume energy~\cite{vRH,vRDH,Graf,DL}. 
Bulk pressure and elastic constants are expected to be similarly sensitive.
Furthermore, our expression for $E_{\rm o}$ is consistent with that obtained 
by van Roij {\it et al}~\cite{vRDH} from an alternative density-functional 
approach. The sole distinction is that our expression, which involves the 
effective counterion density $n_{\rm c}$, incorporates excluded volume
corrections, at least in an approximate fashion.  Although not likely of 
significant consequence at the small volume fractions considered in 
Refs.~\cite{vRH} and \cite{vRDH}, such effects may become important 
at higher concentrations~\cite{DL}.

In conclusion, the linear response approach of Silbert 
{\it et al}~\cite{Silbert1,Silbert2} offers a powerful tool for 
investigating effective electrostatic interactions in charge-stabilized 
colloidal suspensions.
As demonstrated here, a consistent extension to finite-sized macroions 
leads directly to (1) an effective pair interaction between pseudo-macroions 
having precisely the DLVO screened-Coulomb form, but with a modified inverse 
screening length that incorporates excluded volume corrections, and
(2) a density-dependent volume energy that can make a significant contribution 
to the total free energy of salt-free suspensions.
By including the next higher-order response function 
[$\chi^{(2)}$ in Eq.~(\ref{rhoc})], 
the approach can be straightforwardly generalized to include {\it nonlinear} 
response of microions and thereby used to assess the importance of effective 
three-body interactions~\cite{abinitio}.  This is equivalent to
approximating the counterion free energy [Eq.~(\ref{Fc3})] in 
perturbation theory to third order in the macroion-counterion interaction.
An expression for an effective 
triplet interaction already has been derived from a density-functional 
approach~\cite{triplet}.  It remains, however, to analyse the corrections 
that are entailed both to the volume energy {\it and} to the effective 
pair interaction, and to explore the implications for thermodynamic properties.
Experience from {\it ab initio} simulations~\cite{abinitio} and from
the realm of metals~\cite{Louis} suggests that many-body effects become 
significant at sufficiently high densities.
Further outstanding issues are whether in bulk the effective 
pair interaction always retains its screened-Coulomb form, and whether near 
a boundary the interaction can ever become attractive~\cite{Goulding}.
It is hoped that in future the response approach may help to resolve 
these important issues.

\ack
It is a pleasure to thank Anne M.~Denton, Hartmut L\"owen, Hartmut Graf, 
and Christos N.~Likos for helpful discussions.

\end{document}